# The Relationship between Insulin Resistance and Neutrophil to Lymphocyte Ratio

Alicia Shin
1. Torrey Pines High School, 3710 Del Mar Heights Rd, San Diego, CA 92130, United States

alicia.hyesuh.shin@gmail.com

## Abstract

There is increasing interest in the role of chronic inflammation on pathogenesis of various diseases, and one of its markers, high NLR is associated with various mortality and morbidity risk, suggesting IR might be one potential associate factor. However, epidemiological studies on the association between NLR and IR are scarce, and they only included diabetes mellitus patients, excluding the general population. This study aims to determine if there is a direct correlation between NLR and IR in the US general population. The HOMA-IR value was calculated to evaluate the IR of the 3,307 samples provided by the NHANES. As insulin usage could result in inaccurate HOMA-IR estimation, we excluded them and ran a subgroup analysis. Relationship was shown when insulin users were included, having a beta coefficient value of 0.010 (95% confidence interval [CI] of 0.003-0.017). However, without insulin users, the beta value decreased to 0.004 (95% CI of -0.006-0.015). The statistical significance wasn't reached when age, sex, and BMI were adjusted for in the multivariate analyses. Therefore, IR might not explain the variation of NLR value in healthy people, and further studies are needed to reveal the associated factor of high NLR.

## Introduction

There has been an increase in the study of Neutrophil-Lymphocyte Ratio (NLR) and its correlation to various health issues among the human population. NLR is a new biomarker of chronic inflammation which is calculated by dividing the Neutrophil count by the lymphocyte count in blood (Song et al. 2021). Neutrophil is a type of White blood cell (WBC) that supports the healing process of damaged tissues and the process of resolving infections. Their role is to recognize phagocytose microbes, to kill pathogens through mechanisms that include the producing reactive oxygen species, releasing antimicrobial peptides, and expulsing their nuclear content to form traps (Mayadas, Cullere, and Lowell 2014). Lymphocytes are also a type of WBC for the immune system, including the B-Cells and the T-Cells. NLR, therefore, reflects the balance of acute and chronic inflammation (neutrophil count), and adaptive immunity (lymphocyte count) (Song et al. 2021). It is being recognized as a popular marker for medical examinations because NLR calculations are simpler and cheaper compared to other biomarkers (Lou et al. 2015). Previous studies have found out that an increase in NLR is associated with an increase in mortality (Liu et al. 2020), diabetes mellitus (Guo et al. 2015), ischemic stroke (Liu et al. 2020), cerebral hemorrhage (Lattanzi et al. 2019), major cardiac events (Park et al. 2018), sepsis, and infectious diseases (de Jager et al. 2010).

Insulin resistance (IR) is regarded as key pathophysiology of diabetes mellitus, which is when the target tissue of the insulin does not respond to the stimulation (Freeman and Pennings 2021), prohibiting blood glucose levels to decrease. In normal circumstances, as the glucose level increases in the bloodstream, the pancreas releases insulin to stimulate a series of receptors to muscle cells, enabling glucose to enter the cells. Through this process, glucose is converted to a form of energy that is reserved in long-term storage. IR, which prevents this process from happening, is also known to be associated with inflammation (McDade 2012), obesity (Shoelson, Lee, and Goldfine 2006), CVD (Freeman and Pennings 2021, Shoelson, Lee, and Goldfine 2006), nonalcoholic fatty liver disease (Freeman and Pennings 2021), metabolic syndrome (Freeman and Pennings 2021), and polycystic ovary syndrome (Freeman and Pennings 2021).

Several studies suggested linkage between NLR and metabolic syndrome (Surendar et al. 2016, Lou et al. 2015). However, while NLR is highly correlated with metabolic syndrome, it is a cluster of risk factors, and does not directly mean insulin resistance. In one previous study with diabetes patients, IR and NLR have been associated with each other [3]. The two comparison groups they had were DM patients without IR and DM patients with IR. Patients without IR showed lower NLR values than patients with IR, and the logistic regression analysis also revealed significance in the correlation of IR and NLR (Lou et al. 2015). As a result, the study concluded that NLR can play a role as an important biomarker when predicting IR in diabetic patients (Lou et al. 2015). A limitation from this study was that they did not consider insulin use, which might affect the calculation of HOMA-IR, the indicator of insulin resistance (Wallace, Levy, and Matthews 2004).

To current knowledge, there is no study which investigated the relationship between NLR and IR in the general population and with the consideration of insulin use. A new study consisting of the general population and examining the specific DM and insulin usage would be essential to discover the correlation of IR and NLR.

*Hypothesis and Aim*

This study will be examining the association between NLR and IR in the general population, considering DM and insulin treatment status. It was hypothesized that increasing IR will affect the increase of NLR.

## Methods
*Data source*

For the study participants, the general data collected in The National Health and Nutrition Examination Survey (NHANES), a cross-sectional study conducted by the National Center for Health Statistics (NCHS) to obtain a sample that represents the United States population, was used. It has been constantly releasing updated data since 1959 to the public; however, stopped in March 2020 due to the COVID-19 breakout. On the NHANES website, there is information on physical examination, questionnaire data, webinars, and surveys that portray the US population's health and nutrition status in 2-year intervals (Center for Disease

Control and Prevention). The data which is used in this study was obtained through a community-based survey between 2017 and March 2020 Pre-pandemic.

*Study population*

Of the 15,560 people in the data who completed the physical examination between 2017-March 2020, samples over the age of 19 and those who have completed data on insulin usage, BMI, and CBC were included. To establish a more accurate analysis, the following were excluded: participants with a fasting glucose level that exceeds 140 mg/dL, as the normal range of fasting blood glucose level is below 140mg/dL (SHOBHA S. RAO); participants with more than 10,000 WBC in the collected blood sample, because healthy individuals show values between 4,500-10,000 WBC and >10,000 suggest acute inflammatory status (Aminzadeh and Parsa 2011); participants with NLR over 9, as it becomes an outlier considering that normal NLR value in adults is between 0.78 and 3.53, and >9 might mean critically ill status (Rajwani, Cubbon, and Wheatcroft 2012). These factors were eliminated due to the possibility of the data being altered as the factors showed abnormal figures and could be possible outliers. After all the eliminations, 3,375 samples remained.

During the analysis, 68 insulin users were additionally removed as they could be a confounding variable in the overall analysis. This decision was made due to the Homeostasis Model Assessment of Insulin Resistance (HOMA-IR) that is used in this study to measure how much insulin the pancreas creates to regulate blood glucose level. With the inclusion of insulin users, the data on HOMA-IR will show an abnormally large number when compared to individuals who are not on insulin treatment (for detail, see **Figure 4A)**, and show an unsteady insulin concentration in blood depending on when they receive the treatment (Wallace, Levy, and Matthews 2004).

*Data collection*

Final data collection was done on the individuals' gender, race, height, weight, fasting glucose level, BMI, WBC level, and NLR. Body measurements were collected in the Mobile Examination Center (MEC). General information data such as age, gender, and disease status etc., were collected through surveys, and body measurement data were performed by experts with the individual's consent. Overnight fasting blood samples were obtained for fasting glucose, insulin, and other laboratory values.

**Figure 1:** Shows the process of exclusion.

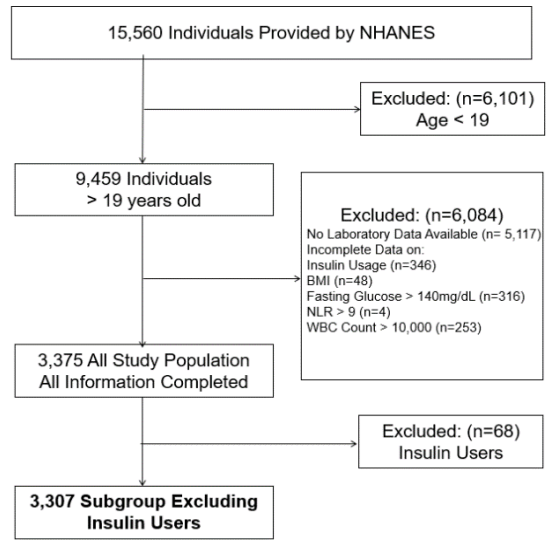

*Statistical analyses*

Descriptive statistics (mean, standard deviation (SD) for continuous variables and number and percentage for categorical variables) were used to describe the study population. The distribution of NLR and HOMA-IR level was visualized by histogram. They showed right skewness for both.

Statistical analyses were performed to see how NLR level related to HOMA-IR and insulin status. The correlation between NRL and HOMA-IR was displayed in a scatter plot and a line of best fit (regression line). Due the right skewness of the NLR and HOMA-IR, the relationship was also displayed after logarithmic transformation, calculating the values of Log HOMA-IR and Log NLR. We also showed it by three groups: non-DM, DM not on insulin, and DM on insulin.

DM and insulin status was considered in our analyses with *a priori* hypothesis. The means (SD) of HOMA-IR and NLR by DM and insulin status was shown by a bar graph. It showed that HOMA-IR and NLR were significantly higher when the subjects were on insulin treatment, therefore, insulin users were excluded in the additional data analysis. Bivariate and Multivariate linear regression analyses were performed to examine statistical significance of the relationship between NLR and HOMA-IR. Multivariate linear regression analyses were performed controlling for age, sex, and BMI, as done in previous study (Surendar et al. 2016), as they could be confounding variables that influence NLR values (Surendar et al. 2016). BMI was controlled because a high BMI is linked to the increase of insulin resistance (Martinez et al. 2017) and NRL (Rajwani, Cubbon, and Wheatcroft 2012) which may also alter the relationship overall. We took HOMA-IR as an independent variable and NLR as a dependent variable, as we hypothesized that insulin resistance would induce chronic inflammation. While bidirectional relationship was suggested in several studies (Lou et al. 2015), the most recent knock-out mice model suggested such a direction (Shimobayashi et al. 2018).

All statistical analyses were made using STATA 14.0 (College station, TX) and

p-values<0.05 were considered statistically significant.

## Results and Discussion
### *Study population*
Table 1 shows the participants characteristics with the sample including insulin users and the sample excluding insulin users. Among the 3,307 participants, 1,731 (52.3%) were female, and mean age was 49.4 (SD 17.8). 298 (8.83%) were diabetes mellitus patients, and 68 (2.01%) used insulin.

**Table 1: Analysis on all participants (N=3,375) and participants without insulin users (N=3,307)**

|  | All participants (N=3,375) | | Excluding insulin user (N= 3,307) | |
|---|---|---|---|---|
|  | N (%) or mean (SD) | | N (%) or mean (SD) | |
| Age | 49.35 | 17.83 | 49.02 | 17.79 |
| Sex, Female | 1762 | 52.21 | 1731.00 | 52.34 |
| Diabetes Mellitus | 298 | 8.83 | 230.00 | 6.95 |
| On insulin | 68 | 2.01 | - | - |
| Hypertension | 1187 | 35.17 | 0.34 | 0.48 |
| Asthma | 518 | 15.35 | 503.00 | 15.21 |
| Arthritis | 945 | 28.00 | 910.00 | 27.52 |
| Heart failure | 99 | 2.93 | 84.00 | 2.54 |
| Coronary heart disease | 119 | 3.53 | 111.00 | 3.36 |
| Stroke | 150 | 4.44 | 140.00 | 4.23 |
| Chronic obstructive pulmonary disease | 267 | 7.91 | 257.00 | 7.77 |
| BMI (kg/m2) | 29.41 | 7.32 | 29.35 | 7.31 |
| Fasting glucose (mg/dL) | 96.60 | 12.69 | 96.34 | 12.33 |
| Fasting insulin (μU/mL) | 13.25 | 17.63 | 12.41 | 11.20 |
| HOMA-IR | 3.30 | 4.73 | 3.06 | 3.05 |
| ALT (IU/L) | 21.81 | 20.56 | 21.9. | 20.73 |
| AST (units/L) | 21.90 | 15.63 | 21.93 | 15.72 |
| ALP (units/L) | 76.41 | 24.60 | 76.17 | 24.37 |
| Albumin (g/dL) | 4.03 | 0.33 | 4.03 | 0.33 |
| Blood urea nitrogen (mg/dL) | 14.54 | 5.57 | 29.35 | 7.31 |
| Sodium Bicarbonate (me) | 25.59 | 2.38 | 25.59 | 2.36 |
| Creatinine | 0.89 | 0.50 | 0.88 | 0.44 |
| Gamma-glutamyl transferase (units/L) | 30.75 | 56.13 | 14.39 | 5.31 |
| Total cholesterol (mg/dL) | 184.37 | 40.31 | 184.89 | 40.25 |

| Triglyceride (mg/dL) | 117.65 | 87.53 | 117.40 | 87.86 |
| Uric acid (mg/dL) | 5.44 | 1.44 | 5.42 | 1.42 |

**Figure 2: Distribution of NLR and HOMA-IR**

    **Figure 2A** shows the distribution of baseline NLR values according to the general population information from NHANES. The graph suggests that the general population (3.307 samples in this study) has a mean of 2 ± 0.02, median of 1.82. Also, the range between the 25th percentile and the 75th percentile was 1.08 for neutrophil and lymphocyte count. **Figure 2B** shows the distribution of baseline HOMA-IR values of the general population. The graph suggests that the mean of 3.3 ± 4.73, median of 2.23. Also, the range between the 25th percentile and the 75th percentile was 2.48 for HOMA-IR.

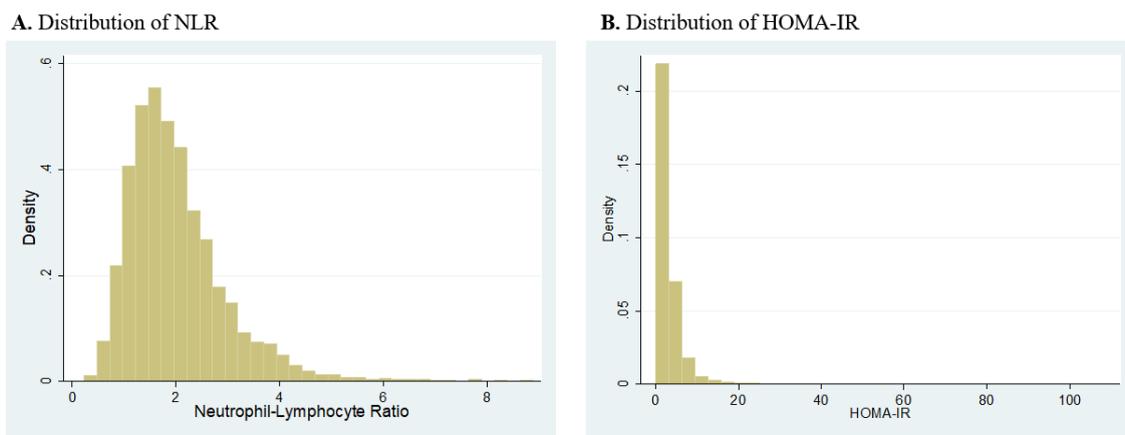

**Figure 2.** The distribution of neutrophil-to-lymphocyte ratios (A) and HOMA-IR (B) among individuals. The x-axis is truncated at an NLR of 9.

**Figure 3: Correlation between NLR and HOMA-IR**

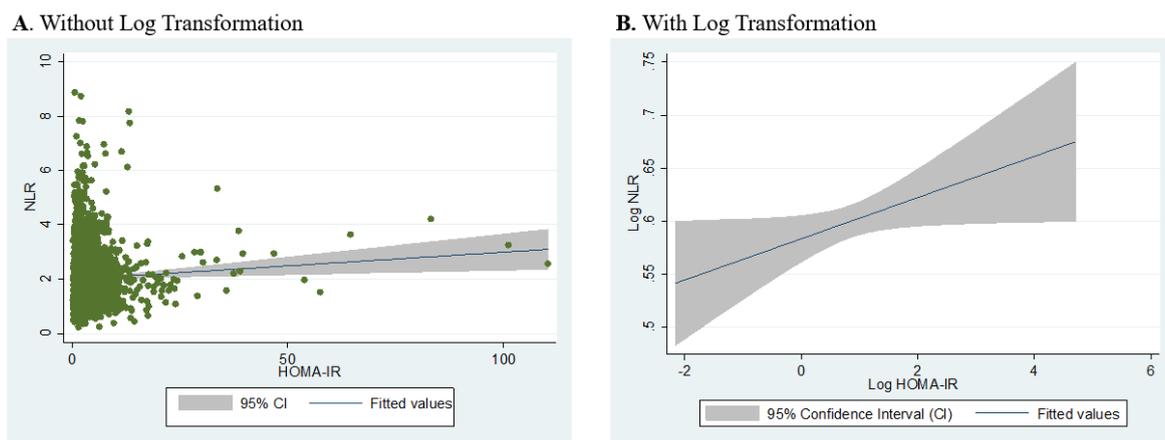

**Figure 3.** The regression line and 95% Confidence Interval of the relationship between Log HOMA-IR and Log NLR before (3A) and after log transformation (3B).

    **Figure 3** presents a weak positive correlation between HOMA-IR and NLR. This graph went through a logarithmic transformation due to the skewness, representing the

relationship between Log HOMA-IR and Log NLR. A weak, positive correlation was shown by the regression line. As the Log HOMA-IR slightly increased, the Log NLR value would also increase.

**Figure 4: Relationship of NLR and HOMA-IR by diabetes and insulin treatment status**

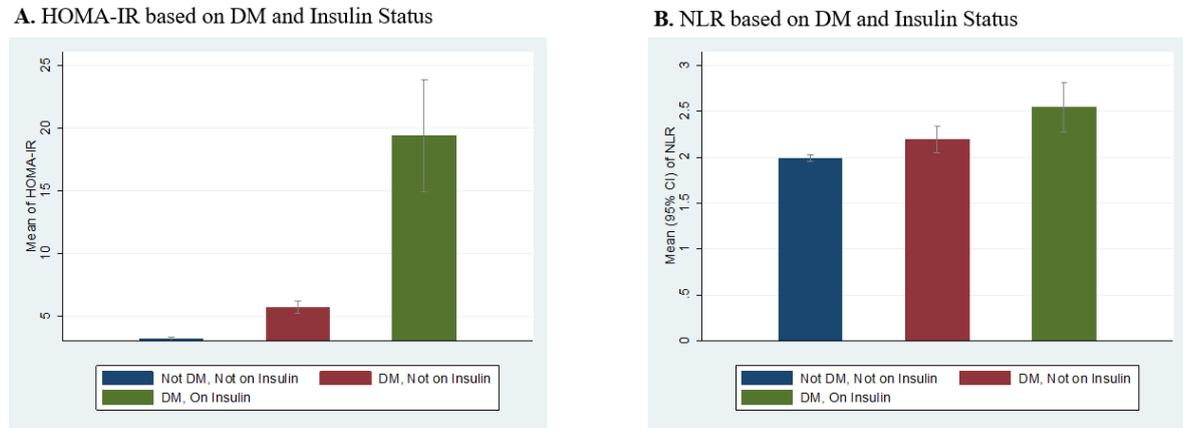

**Figure 4.** Bar graph comparing the mean (95% Confidence Interval) HOMA-IR (A) and NLR level (B) of individuals: not Diabetes Mellitus (DM) and not on insulin, DM and not on insulin, and DM and on insulin.

Figure 4A shows the different mean HOMA-IR value for each criterion. Individuals without DM and insulin had a mean HOMA-IR value of 3.24 ± 0.07, individuals with DM but no insulin treatment had a mean of 5.76 ± 0.27, and individuals with both DM and insulin treatment had a mean of 19.42 ± 2.27. Here, the effect of the insulin treatment on HOMA-IR was clear, suggesting the need to exclude this group in additional analysis.

Figure 4B describes the different NLR value depending on the individual's current DM and insulin status. The results showed the mean 95% confidence interval (CI) for each criterion. Individuals without DM and insulin had a mean NLR value of 2.06 ± 0.02, individuals with DM but no insulin treatment had a mean of 2.32 ± 0.59, and individuals with both DM and insulin treatment had a mean of 2.65 ± 0.10.

**Figure 5: Distribution of NLR depending on DM and Insulin Status**

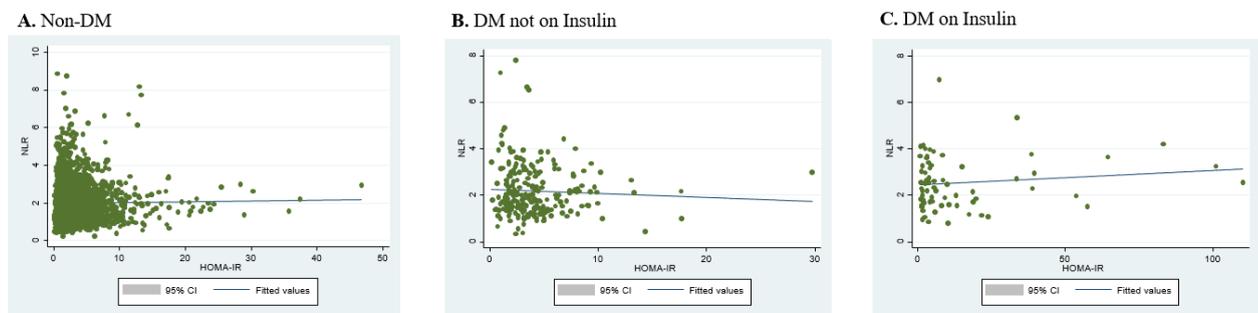

**Figure 5:** Scatterplot for each criterion, comparing HOMA-IR distribution.

**Figure 5** suggests the difference in HOMA-IR - NLR association by DM and insulin status. The non-DM groups showed nearly no significant correlation. Also, DM not on Insulin group showed a weak, negative correlation. However, the DM on the Insulin group showed a drastically increased HOMA-IR value, and therefore, the individuals in the group were eliminated for further analysis.

*Bivariate and Multivariate Linear regression analyses*
**Table 2** shows the difference in the linear regression and significance of all study subjects (n=3,375) and the subjects after excluding insulin users (n=3,307). Bivariate linear regression analysis showed a significant positive relationship between HOMA-IR and NLR (beta-coefficient 0.010, 95% CI 0.003-0.017, P-value 0.004). After adjusting age, sex, and BMI, it lost statistical significance due to the altered values (beta-coefficient 0.004, 95% CI -0.006-0.014).

With the exclusion of insulin users, the beta value changed from 0.01 to 0.004, and the 95% CI interval drastically changed from 0.003-0.017 to -0.006-0.015, indicating that there is no significance. This suggests that there is no true association between HOMA-IR and NLR.

**Table 2: Univariate and Multivariate Linear Regression Model of HOMA-IR on NLR**

| All study subject (N=3,375) | | | | | | |
|---|---|---|---|---|---|---|
| | β | SE | t | P>t | 95% CI | |
| Bivariate | | | | | | |
|   HOMA-IR | 0.010 | 0.003 | 2.880 | 0.004 | 0.003 | 0.017 |
|   Constant | 1.982 | 0.020 | 98.200 | 0.000 | 1.942 | 2.022 |
| Multivariate | | | | | | |
|   Age | 0.010 | 0.001 | 10.670 | 0.000 | 0.008 | 0.012 |
|   Sex | 0.033 | 0.033 | 1.010 | 0.315 | -0.031 | 0.097 |
|   BMI | 0.004 | 0.002 | 1.820 | 0.069 | 0.000 | 0.009 |
|   HOMA-IR | 0.006 | 0.004 | 1.720 | 0.086 | -0.001 | 0.013 |
|   Constant | 1.371 | 0.084 | 16.290 | 0.000 | 1.206 | 1.536 |
| Excluding Insulin Users (N=3,307) | | | | | | |
| Bivariate | | | | | | |
|   HOMA-IR | 0.004 | 0.005 | 0.770 | 0.439 | -0.006 | 0.015 |
|   Constant | 1.992 | 0.024 | 84.510 | 0.000 | 1.945 | 2.038 |
| Multivariate | | | | | | |
|   Age | 0.010 | 0.001 | 10.360 | 0.000 | 0.008 | 0.011 |
|   Sex | 0.028 | 0.033 | 0.850 | 0.396 | -0.037 | 0.093 |
|   BMI | 0.006 | 0.003 | 2.460 | 0.396 | 0.001 | 0.011 |
|   HOMA-IR | -0.003 | 0.006 | -0.550 | 0.584 | -0.015 | 0.009 |
|   Constant | 1.352 | 0.085 | 15.860 | 0.000 | 1.185 | 1.519 |

BMI: Body mass index; HOMA-IR: Homeostasis model assessment of insulin resistance; β: coefficient; SE: standard error; 95% CI: 95% Confidence Interval

*Summary*

In this study using the US general population data, the observation showed a weak, positive association between the NLR and IR in the overall population, but null association when insulin users were excluded.

*Previous Studies/ Mechanism*

We hypothesized that high HOMA-IR value, indicative of insulin resistant state, could predict high NLR because previous preclinical studies suggested a relationship between two. Some studies found out that chronic inflammations can induce systemic insulin resistance: for example, Uysal et al revealed that proinflammatory cytokine TNF-alpha can generate IR by showing protection from obesity-induced insulin resistance in mice lacking TNF-alpha function (Uysal et al. 1997); in addition, Talukdar et al showed neutrophil mediate insulin resistance by using mice without neutrophil elastase, an enzyme that responses to tissue injury, in high fat diet fed mice via secreted elastase (Talukdar et al. 2012).

Others examined the relationship in the opposite direction: Simobayashi used Knockout mice, genetically modified to lack mTORC2 (causing mice to have IR) for observing the relationship between IR and inflammation (Shimobayashi et al. 2018). mTORC2 is a protein complex that regulates glycolysis and the pentose phosphate pathway (Fu and Hall 2020). These mice were also on a high-fat-diet (HFD) and were observed for 10 weeks. Simobayashi found that insulin resistance caused macrophage numbers to increase while B and T cell numbers remained consistent (Shimobayashi et al. 2018). In addition, the increase in macrophages in the KO mice were disproportionate to other variables, indicating that IR promotes inflammation (Shimobayashi et al. 2018).

In line with those studies, an epidemiological study done by Lou et al showed a positive relationship between NLR and IR. They collected data from diabetic patients regardless of their insulin usage status and had an additional 130 healthy subjects for comparison. (Lou et al. 2015). They found out that the NLR values were significantly higher in patients with diabetes than healthy control subjects and interpreted their results as NLR being a marker for IR (Lou et al. 2015) and discovered that T2DM patients were in a state of low-degree chronic inflammation that induces inflammatory factors, elevating neutrophil counts. Luo concluded that NLR has a direct relationship with IR and suggested that NLR can be a biomarker to predict IR in diabetic patients.

However, this study showed results that differ from the above studies because there was no significant correlation between NLR and IR, after controlling for potential confounders and considering insulin use. A difference in Uysal's and Simobayashi's studies was that they used KO mice while this study focused on human subjects. Since the human subjects did not go through a gene deletion process, it could have shown a less definite relationship compared to the correlation shown in mice samples. Also, a difference between Lou's study is that they were focused on patients with diabetes while this study observed the general population. They also included insulin users (insulin-deficient individuals) who dramatically increased the HOMA-IR value. Due to these differences, this study, which focused on the general human population, found no significance or a very weak significance in the correlation of NLR and IR.

## Conclusion

As a result of this study, it was shown that there is no significant association between IR and NLR. Contrary to our hypothesis, insulin resistance did not explain chronic inflammation, represented by NLR. While the increase in each NLR and insulin resistance can cause an increase of mortality, disease, and several types of cancer, they seem to be risk factors which are independent from each other. Therefore, further studies on other factors which increase chronic inflammation and NLR values in the general population, would be needed to develop the strategy to prevent chronic inflammation, which is related to several morbidity and mortality.

**Figure 6: Summary of this study.**

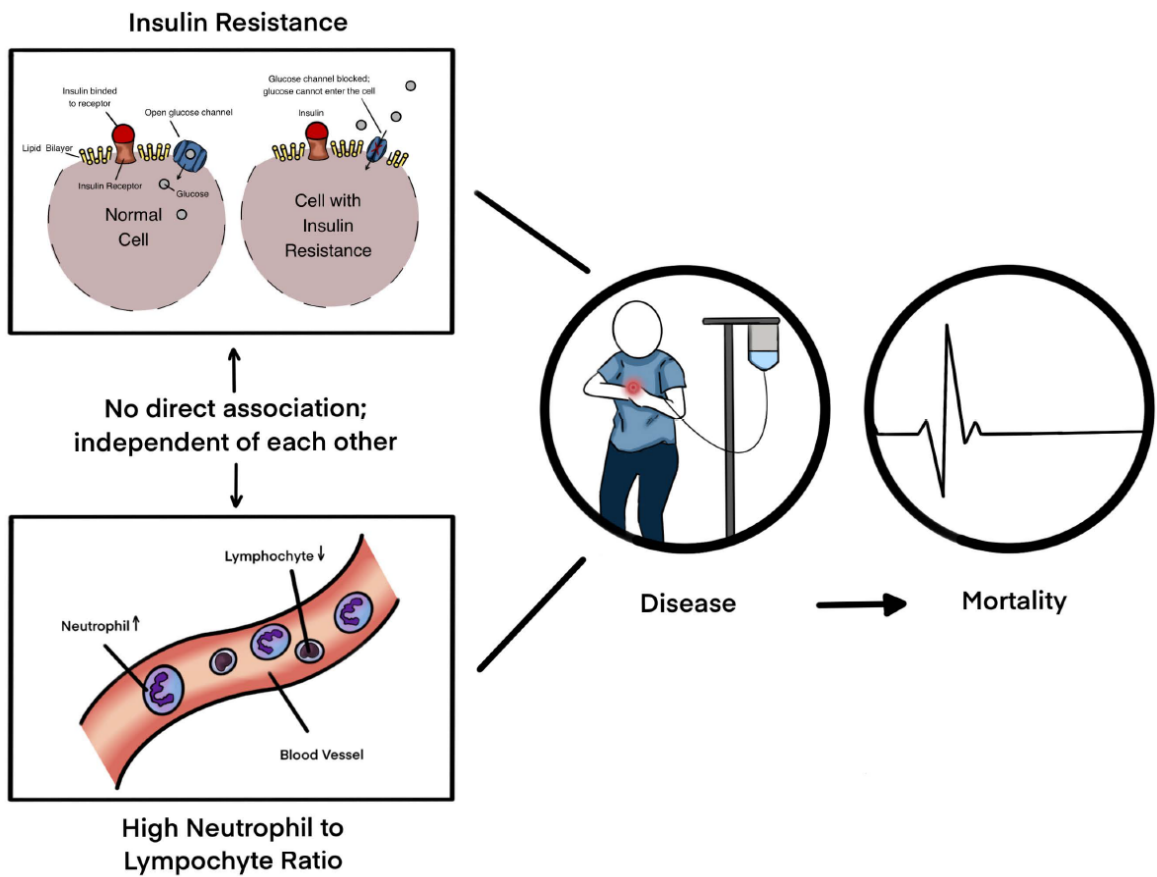

## Acknowledgements
TPHS Medical Illustration Club (Claire Shin)

## Author(s)

Alicia Shin is a junior at Torrey Pines High School, who has been eager to be part of the medical field since a young girl. She hopes to expand her interest in biology and medicine by majoring in biology in college, and possibly studying in medical school.